\documentclass[conference]{IEEEtran}
\IEEEoverridecommandlockouts

\usepackage{cite}
\usepackage{amsmath,amssymb,amsfonts}
\usepackage{algorithmic}
\usepackage{graphicx}
\usepackage{makecell}
\usepackage{textcomp}
\usepackage{xcolor}
\usepackage{subcaption}
\def\BibTeX{{\rm B\kern-.05em{\sc i\kern-.025em b}\kern-.08em
    T\kern-.1667em\lower.7ex\hbox{E}\kern-.125emX}}

\usepackage{booktabs}        
\usepackage{threeparttable}  
\usepackage{siunitx}        
\usepackage[caption=false,font=footnotesize]{subfig}
\usepackage{tikz}
\usepackage{pgfplots}
\pgfplotsset{compat=1.18}
\usetikzlibrary{patterns}
\usepackage{graphicx}   
\usepackage{caption}

\usepackage[table]{xcolor}
\usepackage{colortbl}

\begin{document}

\title{Leveraging Large Language Models to Bridge Cross-Domain Transparency in Stablecoins

}

\author{
Yuexin Xiang$^{1}$,
Yuchen Lei$^{2}$,
Yuanzhe Zhang$^{3}$,
Qin Wang$^{4}$,
Tsz Hon Yuen$^{1}$,
Andreas Deppeler$^{5}$,
Jiangshan Yu$^{6}$
\\[1.5ex]
\IEEEauthorblockA{$^{1}$ Faculty of Information Technology, Monash University, Melbourne, Australia}
\IEEEauthorblockA{$^{2}$ School of Cyber Science and Engineering, Wuhan University, Wuhan, China}
\IEEEauthorblockA{$^{3}$ Digital Trust Centre, Nanyang Technological University, Singapore}
\IEEEauthorblockA{$^{4}$ CSIRO Data61, Sydney, Australia}
\IEEEauthorblockA{$^{5}$ School of Business, Monash University, Bandar Sunway, Malaysia}
\IEEEauthorblockA{$^{6}$ School of Computer Science, The University of Sydney, Sydney, Australia}
}


\maketitle

\begin{abstract}

Stablecoins such as USDT and USDC aspire to peg stability by coupling issuance controls with reserve attestations. In practice, however, transparency remains fragmented across heterogeneous data sources, with key evidence about circulation, reserves, and disclosure dispersed across records that are difficult to connect and interpret jointly. We introduce a large language model (LLM)-based automated framework for bridging cross-domain transparency in stablecoins by aligning issuer disclosures with observable circulation evidence. First, we propose an integrative framework using LLMs to parse documents, extract salient financial indicators, and semantically align reported statements with corresponding market and issuance metrics. Second, we integrate multi-chain issuance records and disclosure documents within a model context protocol (MCP) framework that standardizes LLM access to both quantitative market data and qualitative disclosure narratives. This framework enables unified retrieval and contextual alignment across heterogeneous stablecoin information sources and facilitates consistent analysis. Third, we demonstrate the capability of LLMs to operate across heterogeneous data domains in blockchain analytics, quantifying discrepancies between reported and observed circulation and examining their implications for transparency and price dynamics. Our findings reveal systematic gaps between disclosed and verifiable data, showing that LLM-assisted analysis enhances cross-domain transparency and supports automated, data-driven auditing in decentralized finance (DeFi).
\end{abstract}


\section{Introduction}

Stablecoins play a pivotal role in the digital asset ecosystem by bridging the volatility of cryptocurrencies and the stability of fiat-denominated currencies~\cite {catalini2022some}. They facilitate trading, remittances, and collateralization in decentralized finance (DeFi), and are increasingly regarded as foundational components of the emerging Web3 financial infrastructure~\cite{mai2022stablecoins}. Despite their intended stability, recent crises such as the 2022 collapse of TerraUSD (UST) and LUNA~\cite{ba2025investigating}, as well as the 2025 depegging of Ethena USD (USDe) that occurred during a broader market downturn, which erased nearly 20 billion USD in value, have revealed persistent fragilities in stablecoin mechanisms and limitations in their transparency. These episodes indicate that the resilience of stablecoins depends not only on their structural design but also on the degree of operational transparency maintained by their issuers~\cite{briola2023anatomy,duan2023instability}.

Transparency in this context is fragmented across multiple domains of stablecoin information, including circulation records, issuer attestations, and reserve disclosures. As highlighted in recent studies~\cite{fernandez2024asset}, inconsistencies across these domains can undermine trust and pose systemic risks. Bridging these transparency gaps is therefore essential for assessing the credibility of stablecoins and supporting market integrity.

However, reconciling reported and observable evidence remains challenging. Although aggregate indicators such as total circulating supply are accessible through public data providers, the underlying issuance and redemption activity is distributed across multiple heterogeneous blockchains, which complicates direct verification. At the same time, issuer disclosures such as reserve attestations and financial statements are often released in unstructured formats, including PDF filings and narrative regulatory reports~\cite{fernandez2024asset,catalini2022some}. These records resist automated parsing and often do not align cleanly in timing or granularity with observable circulation data.

Existing research on stablecoin transparency predominantly focuses on observable blockchain activity and related market behavior~\cite{xiang2023babd,tovanich2021empirical,chaudhari2021towards,wahrstatter2023improving,xiang2022leveraging,yu2023predicting}, while issuer disclosures are typically treated as supplementary references rather than analyzable evidence. Despite increasing recognition of transparency as a core determinant of stablecoin credibility~\cite{mahrous2025sok,ante2023systematic}, no existing framework systematically integrates textual issuer disclosures with quantitative issuance evidence. This gap underscores the need for a unified analytical approach that semantically aligns issuer-reported claims with observable circulation indicators.

This fragmentation across heterogeneous information sources highlights a central research gap: the absence of an automated, scalable approach for reconciling observable circulation records with issuer-reported disclosures. To address this gap, this paper introduces an LLM-based automated framework for bridging cross-domain transparency in stablecoins by jointly analyzing quantitative issuance evidence and textual disclosure records. The main contributions are as follows:

\begin{itemize}
\item We propose a novel integrative framework using LLMs to capture, parse, and semantically align heterogeneous stablecoin information sources, extracting key financial indicators from issuer attestations and linking them to corresponding circulation and market metrics.
\item We introduce a model context protocol (MCP) architecture that integrates cross-domain issuance data and issuer disclosures, enabling standardized LLM access, contextual retrieval, and cross-domain reasoning across quantitative and qualitative stablecoin information sources.
\item We demonstrate the ability of LLMs to operate across heterogeneous evidence types in stablecoin analytics, quantifying discrepancies between reported and observed circulation and examining their implications for transparency and price dynamics.
\end{itemize}

In the following section, we review the relevant literature.

\section{Related Work}

\subsection{Transparency and Disclosure in Stablecoins}
Transparency is central to the credibility of stablecoins, as it determines how users and regulators assess the integrity of reserve management and issuance practices. Prior studies have examined transparency from regulatory, financial, and operational perspectives. Fernández et al.~\cite{fernandez2024asset} identify major risks arising from inconsistent attestations and non-standard auditing methods, while Catalini et al.~\cite{catalini2022some} stress that credible reserve disclosures and third-party audits are key to sustaining investor confidence. Ito et al.~\cite{ito2020stablecoin} and Moin et al.~\cite{moin2020sok} provide taxonomies of stablecoin mechanisms, emphasizing the role of disclosure and intervention layers in maintaining trust. Collectively, these works highlight that transparency must be measurable, standardized, and independently verifiable to ensure systemic credibility.

Despite these insights, transparency practices across stablecoin issuers remain fragmented and inconsistent. Attestations are released in heterogeneous formats and at irregular intervals, limiting comparability and independent verification~\cite{mahrous2025sok,ante2023systematic}. Baughman et al.~\cite{baughman2022stable} and Gadzinski et al.~\cite{gadzinski2023stablecoins} further argue that transparency interacts closely with stabilization mechanisms, influencing each stablecoin’s resilience. However, most existing studies do not systematically connect issuer-reported disclosures with observable issuance evidence from blockchain records. A more complete assessment therefore requires a unified analytical framework that integrates these heterogeneous information sources and captures both operational and disclosure-related dimensions of transparency.

\subsection{Mechanisms and Stability Dynamics in Stablecoins}
Stablecoin stability has been widely analyzed through economic, structural, and game-theoretical models. Potter et al.~\cite{potter2024drives} demonstrate that equilibrium stability depends on collateral design and redemption mechanisms, while Klages-Mundt et al.~\cite{klages2020stablecoins} distinguish custodial from decentralized models to explain differences in risk exposure. Li et al.~\cite{li2024stablecoin} and Moin et al.~\cite{moin2020sok} provide systematic taxonomies of design architectures and stabilization mechanisms, showing that collateral composition is a key determinant of performance. Lyons and Viswanath-Natraj~\cite{lyons2023keeps} emphasize that efficient arbitrage access improves peg stability, and Pernice~\cite{pernice2021stablecoin} models corrective forces triggered by deviations. These studies collectively suggest that stability arises from design structures and market access rather than short-term shocks.

Empirical and network-based studies complement these theoretical findings by examining real-world dynamics. Duan and Urquhart~\cite{duan2023instability} and Hoang and Baur~\cite{hoang2024stable} show that major stablecoins frequently deviate from the one-dollar peg, exhibiting heterogeneous reversion speeds. Ba et al.~\cite{ba2025investigating} and Briola et al.~\cite{briola2023anatomy} apply graph-based methods to analyze the UST-LUNA collapse, revealing contagion pathways and systemic fragility across chains. Thanh et al.~\cite{thanh2023stabilities} find that instability in USDT and USDC spills over to other coins, while Ante et al.~\cite{ante2021influence} identify issuance events as drivers of short-term market efficiency. Fernández-Mejia~\cite{fernandez2024extremely} further demonstrates that tail-risk behavior in stablecoins is driven by liquidity and design heterogeneity. Although these studies deepen understanding of stability dynamics, they often treat transparency as an external condition rather than an analyzable component of stability itself.

\subsection{Analysis of Blockchain Records and Issuer Disclosures}

Blockchain-based analysis provides rich observable evidence for studying digital asset ecosystems. Advances in graph analytics and behavioral modeling have enabled fine-grained tracking of transaction activity and systemic dependencies. Xiang et al.~\cite{xiang2023babd} present a large-scale behavioral dataset for Bitcoin addresses, while Tovanich et al.~\cite{tovanich2021empirical} and Chaudhari et al.~\cite{chaudhari2021towards} apply network and anomaly detection methods to identify mining and malicious behaviors. Other works improve transaction classification and community detection through graph-based and subgraph-structural approaches~\cite{wahrstatter2023improving,xiang2022leveraging}. Yu et al.~\cite{yu2023predicting} extend this paradigm using graph neural networks for Web3 asset prediction, demonstrating the scalability of learning-driven blockchain analytics.

A complementary line of work examines document-based evidence such as financial disclosures, audit reports, and regulatory filings. Fernández et al.~\cite{fernandez2024asset} and Catalini et al.~\cite{catalini2022some} highlight that these sources are often unstructured, non-standardized, and difficult to compare across issuers. As a result, many assessments still rely on manual interpretation, introducing ambiguity and time lags. Existing blockchain research primarily targets quantitative transactional and circulation metrics, while the systematic integration of textual issuer disclosures remains underexplored. Combining language-based analysis with blockchain analytics could therefore enable cross-domain methodologies that align reported claims with observable evidence, providing a stronger foundation for automated transparency auditing in digital finance.

\section{Proposed Scheme}

\begin{figure*}[!htbp]
    \centering
    \includegraphics[width=1.8\columnwidth]{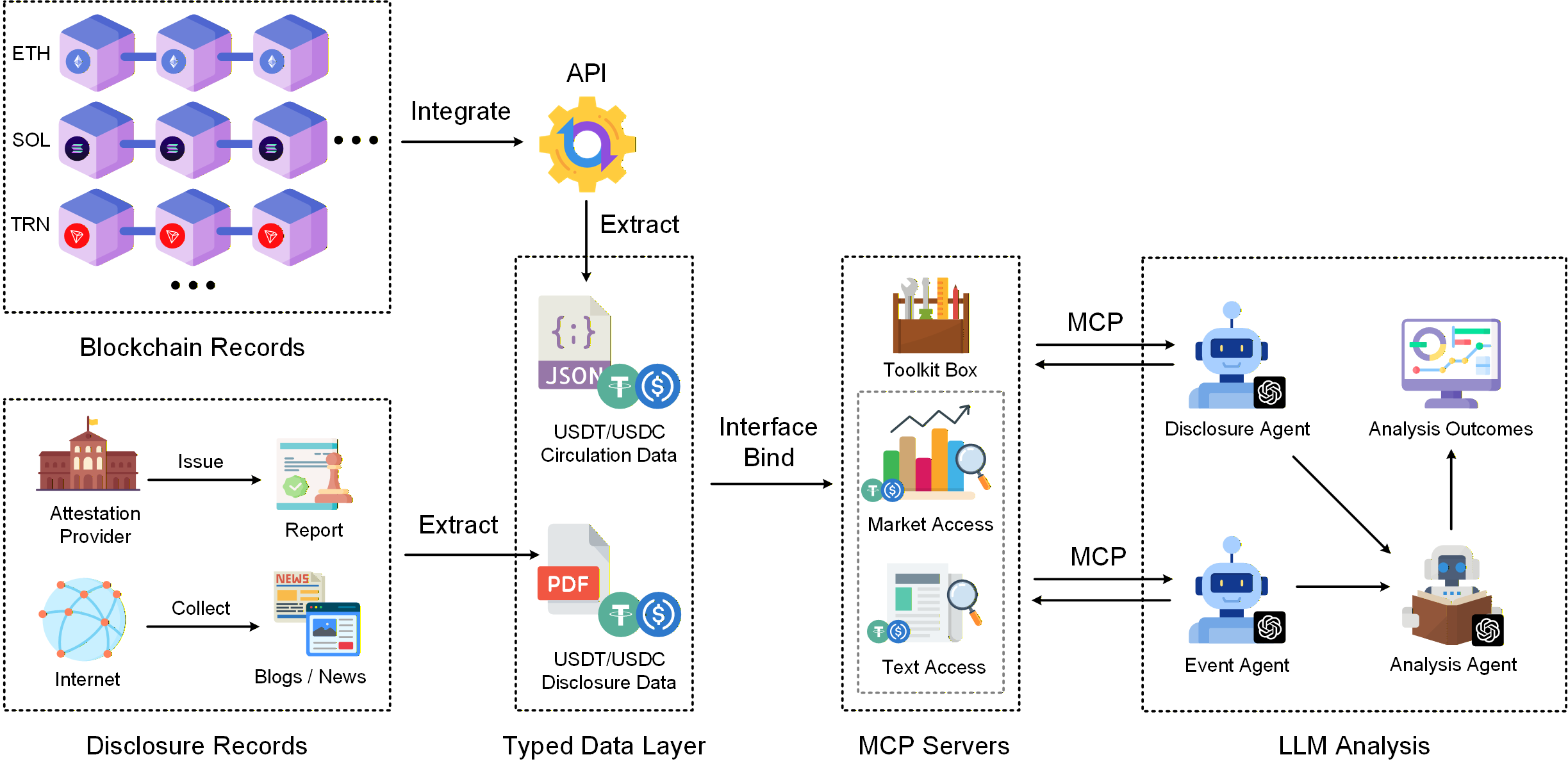}
    \caption{Framework of the proposed scheme.}
    \label{fig:framework}
\end{figure*}

We design an integrated framework that enables time-aligned cross-domain analysis of stablecoins by combining dated circulation records with issuer disclosures. As shown in Fig.~\ref{fig:framework}, the framework is organized into three coordinated modules, each responsible for a distinct stage of the analytical process:
\begin{itemize}
    \item \textbf{Module I -- Data Collection.}  
    Aggregates heterogeneous stablecoin information sources, including dated circulation indicators from \emph{CoinGecko} and issuer disclosures from \emph{Tether} and \emph{Circle}.
    
    \item \textbf{Module II -- MCP Unification.}  
    Standardizes access to these heterogeneous sources via the MCP, aligning all records along a common temporal axis for consistent comparison.

    \item \textbf{Module III -- LLM Analytical Synthesis.}  
    Employs GPT-5 as the analytical core to retrieve, interpret, and reconcile circulation and disclosure evidence, producing structured cross-domain analyses.
\end{itemize}

This modular design enables systematic and interpretable analyses that connect issuer transparency statements with independent market observations, providing a unified and verifiable view of stablecoin integrity.

\subsection{Module I: Data Collection}

This module focuses on integrating quantitative and qualitative dimensions of stablecoin activity through a unified analytical framework. It encompasses data from January 2022 to January 2024, a period sufficient to capture significant market trends and major events influencing the stability and transparency of leading USD-backed stablecoins.

\smallskip
\noindent\textbf{Circulation Data.} 
We collect quantitative circulation indicators such as market capitalization and token price for both USDT and USDC from \emph{CoinGecko}~\cite{coingecko}. These data provide daily aggregated snapshots across supported blockchains, allowing us to track supply fluctuations and price stability over time.

\smallskip
\noindent\textbf{Issuer Disclosures.} 
Complementary disclosure evidence is drawn from the official transparency portals of the two largest stablecoin issuers, \emph{Tether}~\cite{tether_transparency} and \emph{Circle}~\cite{circle_transparency}. We further compile textual disclosures, reserve attestations, and other public statements released by these issuers, complemented by publicly available reports and media sources. These unstructured documents form the basis for semantic extraction and temporal alignment with quantitative circulation indicators.

\subsection{Module II: MCP Module}

\noindent\textbf{Architecture.}  
The MCP provides a unified and auditable interface for accessing heterogeneous stablecoin data, connecting market indicators, issuer disclosures, and textual archives through standardized, typed endpoints. All outputs are timestamped and normalized under a common temporal schema, enabling reproducible cross-modal analysis and eliminating inconsistencies arising from asynchronous reporting.  

The MCP defines two functional classes encompassing five tools that jointly constitute the data-access layer. The \emph{market class} supplies quantitative inputs via daily price snapshots for USDT and USDC, each returning a structured \texttt{PriceSnapshot} object containing price, volume, and capitalization fields. The \emph{textual class} provides qualitative context through three endpoints: two range-query tools that summarize media coverage within date windows, and one direct-access tool retrieving full articles by canonical uniform resource locator (URL).  

All endpoints are schema-defined and stateless, producing deterministic outputs keyed by date or URL so that identical queries always yield identical records. Together, these components ensure consistent alignment between numerical and narrative data streams and support verifiable, reproducible analyses across experiments.

\smallskip
\noindent\textbf{Design Properties.}  
The MCP ensures reliable and reproducible access to heterogeneous stablecoin data through three interrelated properties that jointly guarantee deterministic behavior and modular extensibility. Each market or disclosure endpoint is stateless and keyed by immutable identifiers such as attestation date and blockchain snapshot, ensuring that identical queries always reproduce the same circulation, reserve, or price data. Range queries return compact summaries of daily market activity, supporting efficient and low-latency cross-period analysis.  

Furthermore, market, issuer, and textual channels are modularly separated from the analytical reasoning layer. The LLM accesses stablecoin metrics exclusively through typed MCP calls, removing uncontrolled network dependencies and preserving a verifiable trace of every reserve-to-market comparison. This modularity also allows new issuers or data modalities, such as regulatory attestations, to be integrated declaratively without altering the analytical core. Through these mechanisms, the MCP functions as a transparent and auditable bridge that links market indicators, reserve attestations, and textual disclosures within a consistent temporal and structural framework.

\subsection{Module III: LLM Analytical Synthesis}

\noindent\textbf{Model Selection.}  
The final module employs GPT-5, a state-of-the-art LLM, as the analytical reasoning engine. GPT-5 is selected for its extensive context capacity, reasoning coherence, and robust integration with external computational tools. Its 400,000-token context window allows entire attestation reports, transaction logs, and historical reserve disclosures to be processed jointly, preserving temporal and semantic continuity that earlier models could not maintain. Through lightweight Python interactions with the MCP server, GPT-5 retrieves and integrates observable circulation signals, reported reserves, and market indicators within a unified analytical environment.  

These capabilities enable the model to synthesize heterogeneous information into internally consistent assessments. It evaluates the correspondence between issuer-reported reserves and blockchain-observed circulation, interprets timing differences between disclosure and minting activity, and contextualizes methodological variations across issuers. The resulting analyses provide coherent and interpretable summaries, demonstrating the feasibility of applying a high-capacity reasoning model to transparent and automated evaluation of stablecoin data.

\smallskip
\noindent\textbf{Operational Framework.}  
The analytical reasoning layer implements a \emph{prompt-conditioned code execution framework} in which the LLM functions as a structured decision engine. It orchestrates deterministic data retrieval and synthesis across heterogeneous modalities such as market price feeds and attestation texts through a controlled four-stage reasoning cycle: \texttt{Thought} $\rightarrow$ \texttt{Code} $\rightarrow$ \texttt{Observation} $\rightarrow$ \texttt{Finalize}.  

Guided by a fixed system prompt, the model formulates procedural plans, executes corresponding analytical code, and integrates intermediate observations into successive reasoning steps. The cycle concludes with a finalizing stage that yields structured outputs such as reserve discrepancies, correlation metrics, or alignment summaries between blockchain and disclosure data. Within this operational logic, the model repeatedly applies canonical analytical patterns that formalize how heterogeneous information is interpreted and reconciled.

\smallskip
\noindent\textbf{Agent Design.}  
This multi-agent structure operationalizes the LLM reasoning process, transforming abstract analytical logic into executable and auditable procedures. The architecture consists of three cooperative agents that interact through the MCP interface (as shown in Fig.~\ref{fig:framework}), each fulfilling a distinct role within the reasoning cycle:  

\begin{itemize}
    \item \emph{Disclosure Agent} (\texttt{agent\_disclosure}): Parses issuer reports and attestations to extract structured reserve and liability indicators. It interfaces directly with the MCP text and document modules, filters non-extractable or image-only files, and normalizes all valid numerical fields.  
    
    \item \emph{Event Agent} (\texttt{agent\_event}): Retrieves contemporaneous market and chain-level data from MCP market endpoints, constructing a seven-day observation window that spans three days before and after the disclosure date. It establishes a consistent temporal and quantitative context for subsequent evaluation.  
    
    \item \emph{Analysis Agent} (\texttt{agent\_analysis}): Integrates the structured outputs from the disclosure and event agents to produce the final analytical synthesis. Guided by the system prompt, it ranks indicators by importance, evaluates anomaly persistence, magnitude, and scope, and issues a justified classification within normal, suspicious, and abnormal categories.  
\end{itemize}

All agents operate under a shared prompt-conditioned control schema that enforces deterministic data flow and standardized output formatting. The reasoning process follows a reproducible pathway consisting of disclosure and event retrieval, cross-agent binding through MCP, and integrated analysis and classification. This modular design ensures transparency, interpretability, and auditability of analytical outcomes while allowing the LLM to act as a supervisory controller coordinating specialized data agents via the MCP service interface.

\smallskip
\noindent\textbf{Variable Definitions.}
Table~\ref{tab:variables} summarizes the key variables that jointly constitute the data foundation for consistency assessment. Our analysis integrates market data, issuer disclosures, and derived alignment metrics to evaluate the coherence between observed circulation and reported reserves. This unified framework enables systematic comparison between observable circulation signals and reported reserves, forming the basis for quantitative transparency analysis.

In addition to the variables summarized in Table~\ref{tab:variables}, the analysis produces an aggregate result referred to as \texttt{analysis\_outcome}. This variable categorizes each observation into three levels: normal, suspicious, and abnormal, based on rule-based thresholds for reserve coverage, attestation quality, and market consistency indicators. Observations with reserve deficits, stale or proxy attestations, or large market discrepancies are labeled as \emph{abnormal}. Cases showing moderate deviations or elevated trading volatility are marked as \emph{suspicious}, while all others are considered \emph{normal}, indicating alignment between reported reserves and observed market activity.

\begin{table}[h!]
\centering
\caption{Key variables for stablecoin data analysis}
\label{tab:variables}
\scriptsize
\renewcommand{\arraystretch}{1.3}
\begin{tabular}{lp{5.75cm}}
\toprule
\textbf{Variable} & \textbf{Description} \\
\midrule
\texttt{report\_date} & Reference date of the attestation report. \\
\texttt{price\_usd} & Market price in US dollars. \\
\texttt{mcap\_usd} & Market capitalization derived from supply and price. \\
\texttt{volume\_daily} & Trading volume over the day. \\
\texttt{turnover\_ratio} & Liquidity ratio of volume to market capitalization. \\
\texttt{peg\_deviation} & Deviation of the market price from the one-dollar peg. \\
\texttt{volatility\_daily} & Short-term price volatility over a daily window. \\
\texttt{circulation\_rep} & Reported circulation held by external holders. \\
\texttt{asset\_value} & Total fair value of reserve assets. \\
\texttt{liability\_value} & Total liabilities reported to token holders. \\
\texttt{coverage\_ratio} & Ratio of total assets to total liabilities. \\
\texttt{implied\_mcap} & Market capitalization implied by attested supply and price. \\
\texttt{supply\_gap} & Gap between reported and observed circulation. \\
\bottomrule
\end{tabular}
\end{table}

In the next section, we demonstrate the practical performance of this approach by applying it to USDT and USDC analysis, quantifying reserve discrepancies, assessing their correlation with market prices, and comparing cross-asset transparency patterns.

\section{Empirical Results}

\begin{table*}[!htbp]
\centering
\normalsize
\caption{USDT data overview and analysis outcomes}
\label{tab:usdt-main}
\resizebox{\textwidth}{!}{%
  \renewcommand{\arraystretch}{1.5}
\begin{tabular}{c|rrrrrrrrrrrr|c}
\cmidrule{3-14}
\multicolumn{1}{c}{\textbf{report\_date}} & 
\rotatebox{30}{\textbf{price\_usd}} & 
\rotatebox{30}{\textbf{mcap\_usd}} & 
\rotatebox{30}{\textbf{volume\_daily}} & 
\rotatebox{30}{\textbf{turnover\_ratio}} &
\rotatebox{30}{\makecell{\textbf{peg\_deviation}\\ (\%)}} & 
\rotatebox{30}{\textbf{volatility\_daily}} & 
\rotatebox{30}{\textbf{circulation\_rep}} & 
\rotatebox{30}{\textbf{asset\_value}} & 
\rotatebox{30}{\textbf{liability\_value}} & 
\rotatebox{30}{\textbf{coverage\_ratio}} & 
\rotatebox{30}{\textbf{implied\_mcap}} & 
\multicolumn{1}{c}{\rotatebox{30}{\makecell{\textbf{supply\_gap }\\(\%)}}} & 
\rotatebox{30}{\makecell{\textbf{analysis\_}\\ \textbf{outcome}}} \\
\midrule
18/05/2022 & 9.996E-01 & 8.227E+10 & 6.479E+10 & 7.875E-01 & -3.550E-02 & 6.349E-02 & 8.219E+10 & 8.242E+10 & 8.226E+10 & 1.002E+00 & 8.216E+10 & 1.384E-01 & \cellcolor{red!25} abnormal \\
10/08/2022 & 1.001E+00 & 6.630E+10 & 3.764E+10 & 5.676E-01 & 1.486E-01 & 3.043E-01 & 6.627E+10 & 6.641E+10 & 6.620E+10 & 1.003E+00 & 6.637E+10 & -1.000E-01 & \cellcolor{yellow!30} suspicious \\
10/11/2022 & 1.002E+00 & 6.785E+10 & 2.380E+10 & 3.508E-01 & 1.976E-01 & 1.074E-01 & 6.781E+10 & 6.806E+10 & 6.781E+10 & 1.004E+00 & 6.794E+10 & -1.270E-01 & \cellcolor{green!30} normal \\
08/02/2023 & 1.000E+00 & 6.622E+10 & 1.727E+10 & 2.607E-01 & 1.789E-02 & 3.847E-02 & 6.606E+10 & 6.704E+10 & 6.608E+10 & 1.015E+00 & 6.607E+10 & 2.334E-01 & \cellcolor{green!30} normal \\
09/05/2023 & 1.002E+00 & 7.979E+10 & 1.894E+10 & 2.373E-01 & 2.022E-01 & 8.336E-02 & 7.945E+10 & 8.183E+10 & 7.939E+10 & 1.031E+00 & 7.961E+10 & 2.263E-01 & \cellcolor{green!30} normal \\
31/07/2023 & 9.990E-01 & 8.328E+10 & 9.642E+09 & 1.158E-01 & -9.810E-02 & 8.073E-02 & 8.326E+10 & 8.650E+10 & 8.320E+10 & 1.040E+00 & 8.318E+10 & 1.152E-01 & \cellcolor{green!30} normal \\
31/10/2023 & 9.999E-01 & 8.327E+10 & 1.842E+10 & 2.213E-01 & -1.380E-02 & 0.000E+00 & 8.324E+10 & 8.638E+10 & 8.318E+10 & 1.039E+00 & 8.323E+10 & 5.153E-02 & \cellcolor{green!30} normal \\
31/01/2024 & 1.002E+00 & 9.172E+10 & 5.521E+10 & 6.020E-01 & 2.024E-01 & 1.114E-01 & 9.166E+10 & 9.702E+10 & 9.160E+10 & 1.059E+00 & 9.185E+10 & -1.440E-01 & \cellcolor{green!30} normal \\
\bottomrule
\end{tabular}
}
\end{table*}

\begin{table*}[!htbp]
\centering
\normalsize
\caption{USDC data overview and analysis outcomes}
\label{tab:usdc-main}
\resizebox{\textwidth}{!}{%
  \renewcommand{\arraystretch}{1.5}
\begin{tabular}{c|rrrrrrrrrrrr|c}
\cmidrule{3-14}
\multicolumn{1}{c}{\textbf{report\_date}} & 
\rotatebox{30}{\textbf{price\_usd}} & 
\rotatebox{30}{\textbf{mcap\_usd}} & 
\rotatebox{30}{\textbf{volume\_daily}} & 
\rotatebox{30}{\textbf{turnover\_ratio}} & 
\rotatebox{30}{\textbf{peg\_deviation (\%)}} & 
\rotatebox{30}{\textbf{volatility\_daily}} & 
\rotatebox{30}{\textbf{circulation\_rep}} & 
\rotatebox{30}{\textbf{asset\_value}} & 
\rotatebox{30}{\textbf{liability\_value}} & 
\rotatebox{30}{\textbf{coverage\_ratio}} & 
\rotatebox{30}{\textbf{implied\_mcap}} & 
\multicolumn{1}{c}{\rotatebox{30}{\textbf{supply\_gap (\%)}}} &
\rotatebox{30}{\makecell{\textbf{analysis\_} \\ \textbf{outcome} }} \\
\midrule
25/02/2022 & 1.001E+00 & 4.979E+10 & 1.917E+09 & 3.850E-02 & 7.130E-02 & 0.000E+00 & 5.003E+10 & 5.003E+10 & 5.003E+10 & 1.000E+00 & 5.007E+10 & -5.450E-01 & \cellcolor{green!20} normal \\
31/03/2022 & 1.002E+00 & 5.357E+10 & 3.635E+09 & 6.786E-02 & 2.436E-01 & 1.438E-02 & 5.350E+10 & 5.350E+10 & 5.350E+10 & 1.000E+00 & 5.363E+10 & -1.070E-01 & \cellcolor{green!20} normal \\
29/04/2022 & 1.003E+00 & 5.185E+10 & 3.853E+09 & 7.430E-02 & 3.268E-01 & 2.310E-01 & 5.139E+10 & 5.139E+10 & 5.139E+10 & 1.000E+00 & 5.156E+10 & 5.750E-01 & \cellcolor{green!20} normal \\
23/05/2022 & 9.984E-01 & 4.914E+10 & 4.521E+09 & 9.200E-02 & -1.640E-01 & 6.748E-02 & 4.926E+10 & 4.926E+10 & 4.926E+10 & 1.000E+00 & 4.918E+10 & -7.750E-02 & \cellcolor{green!20} normal \\
28/07/2022 & 1.001E+00 & 5.581E+10 & 4.017E+09 & 7.198E-02 & 1.360E-01 & 4.323E-02 & 5.557E+10 & 5.557E+10 & 5.557E+10 & 1.000E+00 & 5.565E+10 & 3.006E-01 & \cellcolor{green!20} normal \\
24/08/2022 & 9.988E-01 & 5.440E+10 & 8.372E+09 & 1.539E-01 & -1.200E-01 & 1.845E-01 & 5.449E+10 & 5.462E+10 & 5.449E+10 & 1.002E+00 & 5.442E+10 & -4.180E-02 & \cellcolor{green!20} normal \\
23/09/2022 & 1.000E+00 & 5.184E+10 & 4.652E+09 & 8.973E-02 & 1.376E-02 & 1.655E-01 & 5.226E+10 & 5.243E+10 & 5.226E+10 & 1.003E+00 & 5.227E+10 & -8.140E-01 & \cellcolor{green!20} normal \\
25/10/2022 & 1.000E+00 & 4.743E+10 & 2.504E+09 & 5.280E-02 & 3.755E-02 & 2.172E-01 & 4.726E+10 & 4.748E+10 & 4.726E+10 & 1.005E+00 & 4.728E+10 & 3.165E-01 & \cellcolor{green!20} normal \\
22/11/2022 & 1.000E+00 & 4.258E+10 & 6.534E+09 & 1.534E-01 & -3.630E-03 & 7.794E-02 & 4.351E+10 & 4.375E+10 & 4.351E+10 & 1.006E+00 & 4.351E+10 & -2.120E+00 & \cellcolor{yellow!30} suspicious \\
22/12/2022 & 1.001E+00 & 4.323E+10 & 2.212E+09 & 5.118E-02 & 1.218E-01 & 7.651E-02 & 4.324E+10 & 4.340E+10 & 4.324E+10 & 1.004E+00 & 4.329E+10 & -1.510E-01 & \cellcolor{green!20} normal \\
25/01/2023 & 1.000E+00 & 4.469E+10 & 1.550E+09 & 3.469E-02 & 2.837E-02 & 2.471E-02 & 4.455E+10 & 4.469E+10 & 4.455E+10 & 1.003E+00 & 4.457E+10 & 2.868E-01 & \cellcolor{green!20} normal \\
02/03/2023 & 1.002E+00 & 4.254E+10 & 4.071E+09 & 9.571E-02 & 2.254E-01 & 1.240E-01 & 4.229E+10 & 4.234E+10 & 4.229E+10 & 1.001E+00 & 4.238E+10 & 3.701E-01 & \cellcolor{green!20}  normal \\
30/03/2023 & 1.001E+00 & 4.256E+10 & 3.004E+09 & 7.059E-02 & 7.531E-02 & 5.923E-02 & 4.240E+10 & 4.246E+10 & 4.240E+10 & 1.001E+00 & 4.244E+10 & 2.789E-01 & \cellcolor{green!20} normal \\
28/04/2023 & 1.001E+00 & 3.256E+10 & 2.906E+09 & 8.925E-02 & 1.369E-01 & 8.751E-02 & 3.252E+10 & 3.257E+10 & 3.252E+10 & 1.002E+00 & 3.256E+10 & 4.528E-03 & \cellcolor{green!20} normal \\
30/05/2023 & 1.000E+00 & 3.051E+10 & 4.963E+09 & 1.627E-01 & 4.030E-02 & 5.279E-03 & 3.049E+10 & 3.055E+10 & 3.049E+10 & 1.002E+00 & 3.051E+10 & 2.788E-03 & \cellcolor{green!20} normal \\
30/06/2023 & 9.997E-01 & 2.890E+10 & 2.594E+09 & 8.974E-02 & -3.220E-02 & 1.171E-01 & 2.887E+10 & 2.893E+10 & 2.887E+10 & 1.002E+00 & 2.886E+10 & 1.413E-01 & \cellcolor{green!20} normal \\
28/07/2023 & 9.998E-01 & 2.795E+10 & 3.538E+09 & 1.266E-01 & -2.480E-02 & 1.184E-02 & 2.738E+10 & 2.744E+10 & 2.738E+10 & 1.002E+00 & 2.738E+10 & 2.104E+00 & \cellcolor{yellow!30} suspicious \\
30/08/2023 & 9.997E-01 & 2.619E+10 & 4.492E+09 & 1.715E-01 & -2.620E-02 & 1.721E-01 & 2.641E+10 & 2.646E+10 & 2.641E+10 & 1.002E+00 & 2.640E+10 & -7.830E-01 & \cellcolor{green!20} normal \\
29/09/2023 & 9.998E-01 & 2.614E+10 & 4.230E+09 & 1.618E-01 & -2.360E-02 & 3.723E-04 & 2.615E+10 & 2.621E+10 & 2.615E+10 & 1.002E+00 & 2.615E+10 & -8.150E-03 & \cellcolor{green!20} normal \\
30/10/2023 & 9.996E-01 & 2.558E+10 & 6.656E+09 & 2.602E-01 & -4.140E-02 & 6.145E-04 & 2.497E+10 & 2.503E+10 & 2.497E+10 & 1.002E+00 & 2.496E+10 & 2.469E+00 & \cellcolor{yellow!30} suspicious \\
30/11/2023 & 1.000E+00 & 2.493E+10 & 7.160E+09 & 2.872E-01 & 8.055E-03 & 2.304E-01 & 2.467E+10 & 2.472E+10 & 2.467E+10 & 1.002E+00 & 2.467E+10 & 1.065E+00 & \cellcolor{yellow!30} suspicious \\
22/12/2023 & 1.000E+00 & 2.455E+10 & 6.943E+09 & 2.827E-01 & 2.507E-02 & 4.532E-02 & 2.448E+10 & 2.453E+10 & 2.448E+10 & 1.002E+00 & 2.448E+10 & 2.911E-01 & \cellcolor{green!20} normal \\
30/01/2024 & 9.985E-01 & 2.484E+10 & 1.298E+10 & 5.223E-01 & -1.480E-01 & 1.288E-01 & 2.464E+10 & 2.469E+10 & 2.464E+10 & 1.002E+00 & 2.460E+10 & 9.665E-01 &  \cellcolor{green!20} normal \\
\bottomrule
\end{tabular}
}
\vspace{2mm}
\end{table*}

\subsection{Disclosure Analysis}
\noindent\textbf{Overview.} This subsection examines how issuer attestations and circulation indicators together reveal the reserve adequacy, transparency, and operational behavior of the two dominant fiat-backed stablecoins, USDT and USDC. Each observation aligns a formal attestation snapshot with contemporaneous circulation and price indicators drawn from a seven-day window centered on the report date, enabling a structured comparison between what issuers disclose and what observable stablecoin activity indicates.

The analysis emphasizes how attestation quality, reporting cadence, and reserve composition interact with measurable circulation indicators such as the coverage ratio, turnover ratio, and peg deviation, tracing how reserve information is reflected in observable price and liquidity dynamics. The empirical outcomes summarized in Tables~\ref{tab:usdt-main} and~\ref{tab:usdc-main} provide the quantitative basis for this examination.

\smallskip
\noindent\textbf{USDT.}
As shown in Table~\ref{tab:usdt-main}, USDT maintains a consistently narrow price band around its one dollar target throughout all eight observations, with deviations rarely exceeding around 0.2\%. Market capitalization remains large and relatively stable, hovering between USD 66 billion (abbreviated as \$66B) and \$92B while daily trading volumes reach tens of billions. This implies high liquidity and continuous market participation. The turnover ratio oscillates between 0.1 and 0.8, peaking during mid-2022, which reflects intense trading activity rather than structural reserve stress. Reported reserves exhibit strong nominal coverage where coverage ratios range from 1.00 to 1.06, and reported liabilities closely match observable circulation with supply gaps generally below 0.3\%.  

Despite this overall adequacy, early observations reveal moments of tension between disclosure timing and market conditions. The May 2022 attestation is classified as abnormal, and August 2022 as suspicious, both coinciding with high turnover above 0.55 and elevated market volatility. These classifications indicate that attestation data lagged real-time adjustments in circulation during a period of broader crypto-market stress. Afterward, turnover normalizes below 0.25, peg deviations tighten toward zero, and classifications return to normal. Nonetheless, Tether’s irregular attestation schedule (typically quarterly) and limited auditor transparency constrain the precision of its informational signal.  

Overall, USDT displays strong peg stability and liquidity resilience but weaker disclosure reliability. Its coverage remains ample, yet the irregular frequency and limited assurance level of attestations suggest that market confidence is sustained more by liquidity depth and convertibility than by the clarity or rigor of its reserve reporting.

\smallskip
\noindent\textbf{USDC.}
Table~\ref{tab:usdc-main} illustrates that USDC likewise sustains a tight peg, with prices consistently within around 0.15\% of \$1.00 across 2022--2024. Market capitalization declines from about \$50B in early 2022 to roughly \$25B by late 2023, but this contraction is accompanied by steady coverage ratios near 1.00--1.01 and small supply gaps, indicating accurate synchronization between reported reserves and circulating supply. Trading volumes fluctuate in the low single-digit billions per day, and turnover remains moderate that is mostly between 0.05 and 0.3, suggesting more stable transactional use rather than speculative churn.  

Most USDC disclosures are labeled normal, with a few suspicious designations in November 2022 and again during mid-late 2023. These correspond to short-lived increases in turnover (up to 0.29) and small shifts in reported circulation, not to peg divergence or reserve shortfall. Crucially, each attestation is issued on a fixed monthly basis by Grant Thornton using standardized accounting procedures, which reduces temporal lag and enhances methodological transparency. This consistent cadence ensures that reserve data remain timely and comparable across periods, facilitating market verification and minimizing informational asymmetry.  

In sum, USDC’s disclosure framework combines high transparency with operational regularity. Even amid supply contraction, the alignment between attested and market values and the predictability of reporting cadence indicate strong disclosure governance and a higher level of informational assurance than its peers.

\smallskip
\noindent\textbf{Comparative Insights.}
Both USDT and USDC maintain effective reserve coverage and close peg adherence, yet they diverge systematically along three dimensions: market scale, disclosure practice, and observed market behavior.

\begin{figure}[!htbp]
  \centering
  \begin{subfigure}[b]{0.48\textwidth}
    \centering
    \includegraphics[width=\textwidth]{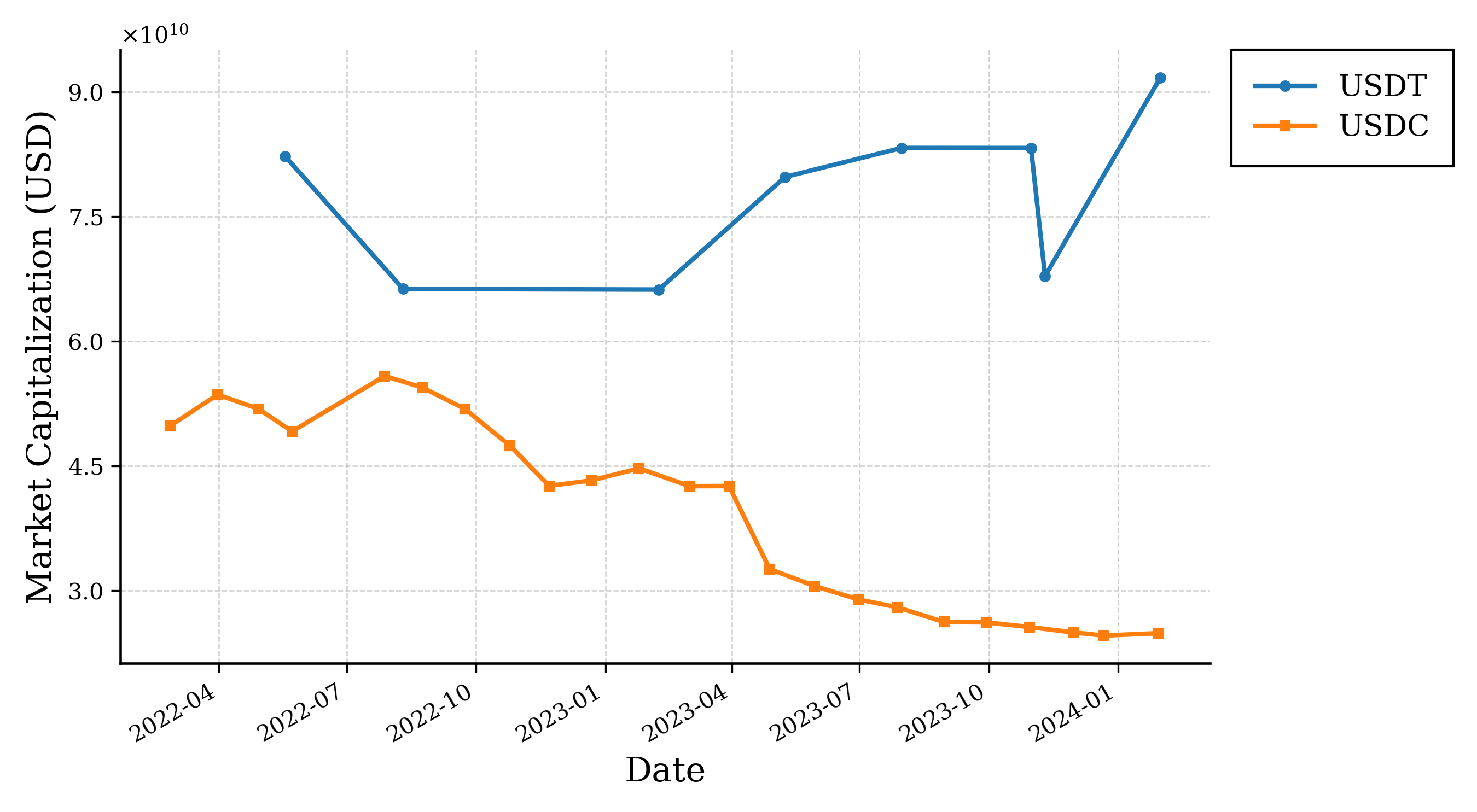}
    \caption{Market capitalization}
  \end{subfigure}
  \hfill
  \begin{subfigure}[b]{0.48\textwidth}
    \centering
    \includegraphics[width=\textwidth]{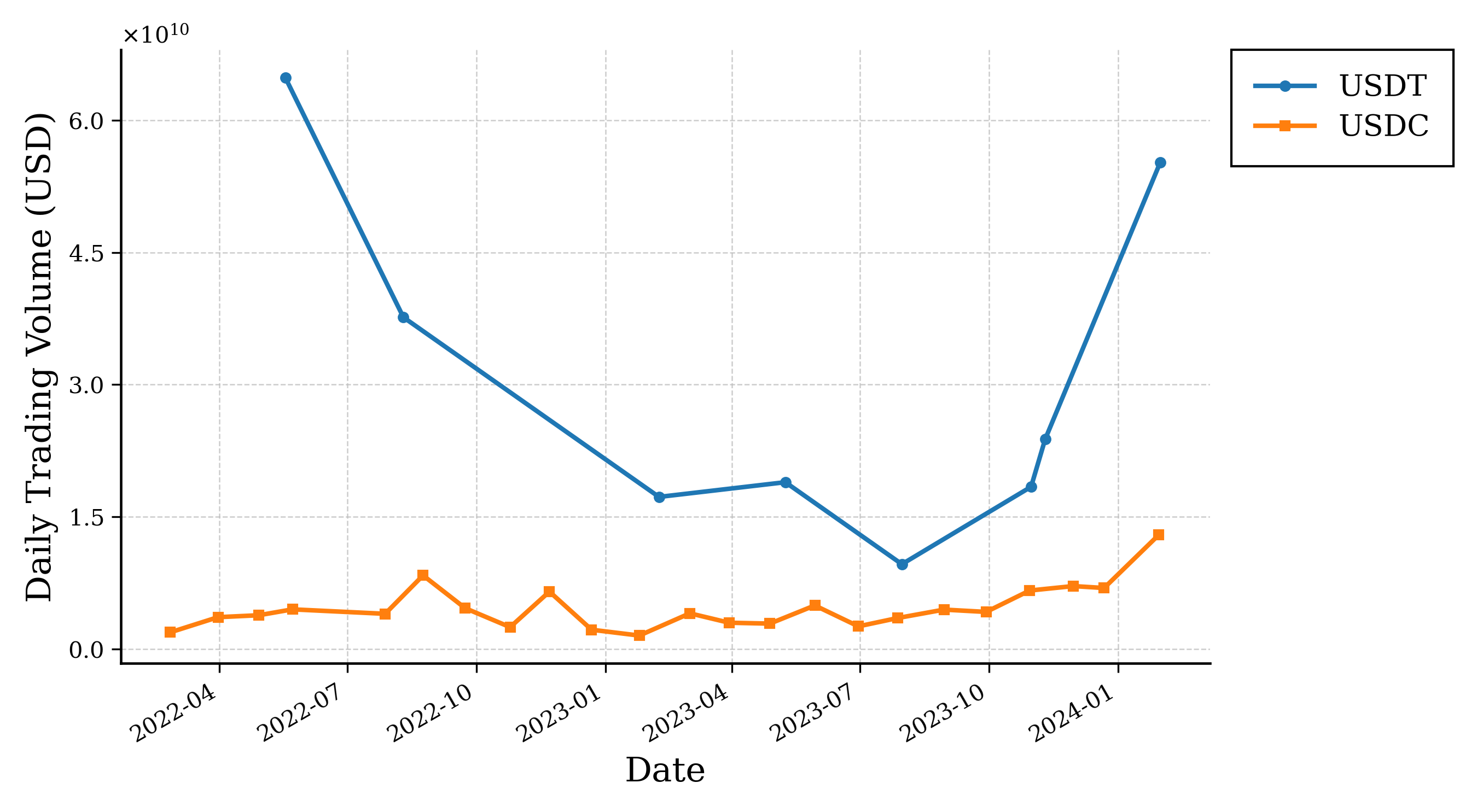}
    \caption{Daily trading volume}
  \end{subfigure}
  \caption{Comparative scale and liquidity indicators of major stablecoins.}
  \label{fig:market_liquidity}
\end{figure}

First, in terms of \emph{scale} and \emph{liquidity}, USDT dominates: its capitalization and daily trading activity are several times larger, supporting its role as the primary settlement medium in crypto markets. As shown in Fig.~\ref{fig:market_liquidity}, USDT’s market capitalization remains above \$60B while USDC steadily declines below \$30B, and its trading volume consistently exceeds that of USDC by a wide margin. These patterns confirm USDT’s dominant market presence and substantiate the interpretation that its scale advantage underpins higher transactional velocity. However, this same liquidity intensity contributes to greater short-term variability in turnover and occasional classification anomalies, particularly during stress periods.

\begin{figure}[!htbp]
  \centering
  \includegraphics[width=0.48\textwidth]{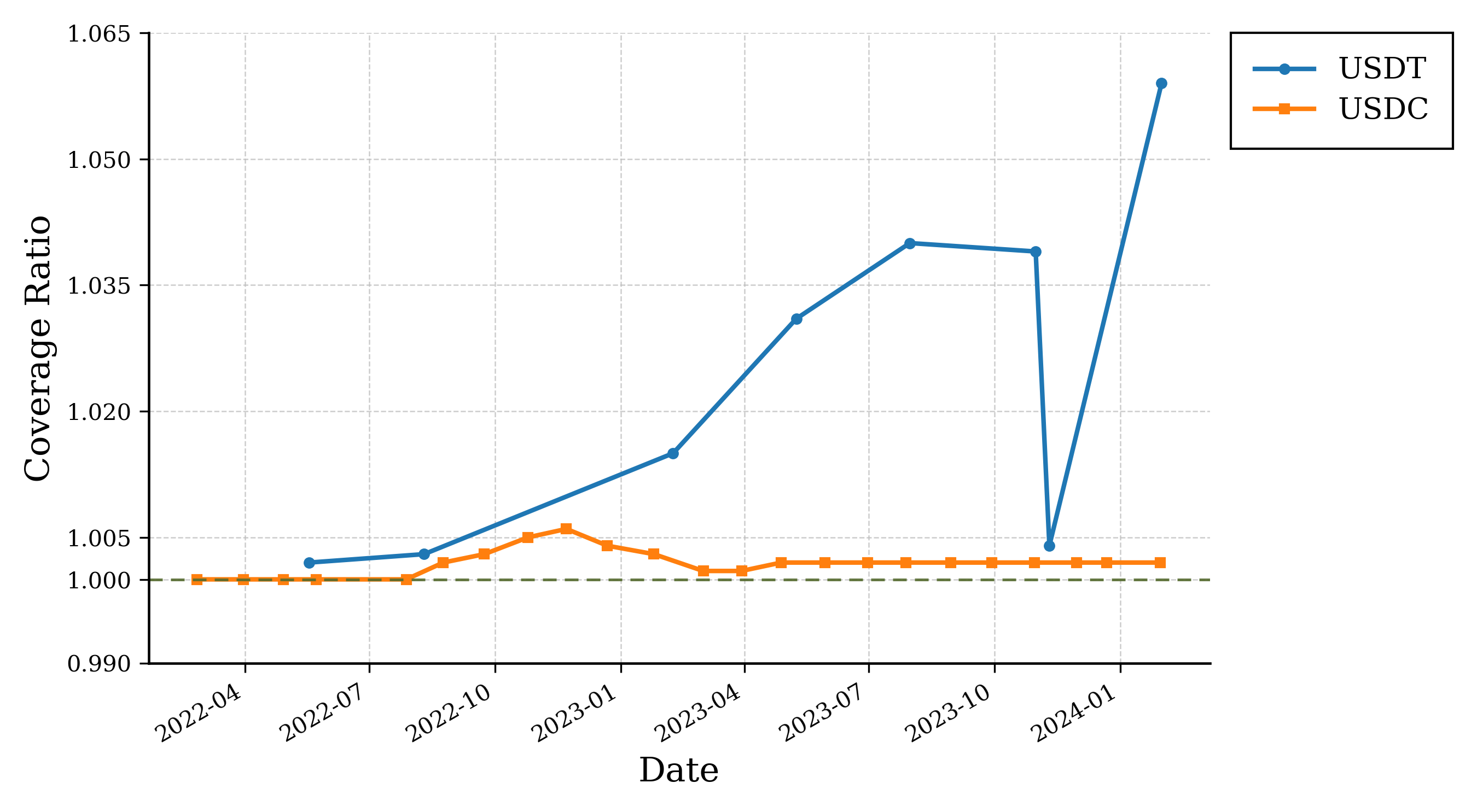}
  \caption{Comparative reserve coverage and disclosure indicators of major stablecoins.}
  \label{fig:coverage_ratio}
\end{figure}

Second, regarding \emph{disclosure cadence} and \emph{assurance}, USDC’s standardized monthly attestations contrast with USDT’s irregular and less formal reporting cycle. As illustrated in Fig.~\ref{fig:coverage_ratio}, both stablecoins maintain coverage ratios at or above one, but USDT’s ratio rises sharply after mid-2022, exceeding 1.05 by early 2024, while USDC stays close to parity throughout. This divergence highlights how USDC’s stable and well-verified disclosures support steady confidence, whereas USDT’s more variable reserve margin reflects less frequent updates and greater sensitivity to market conditions. While both maintain coverage above unity, USDC’s higher attestation frequency and consistent auditor engagement strengthen the reliability of its reserve representation, whereas USDT’s episodic reports leave temporary informational gaps that coincide with market stress episodes.

\begin{figure}[!htbp]
  \centering
  \begin{subfigure}[b]{0.48\textwidth}
    \centering
    \includegraphics[width=\textwidth]{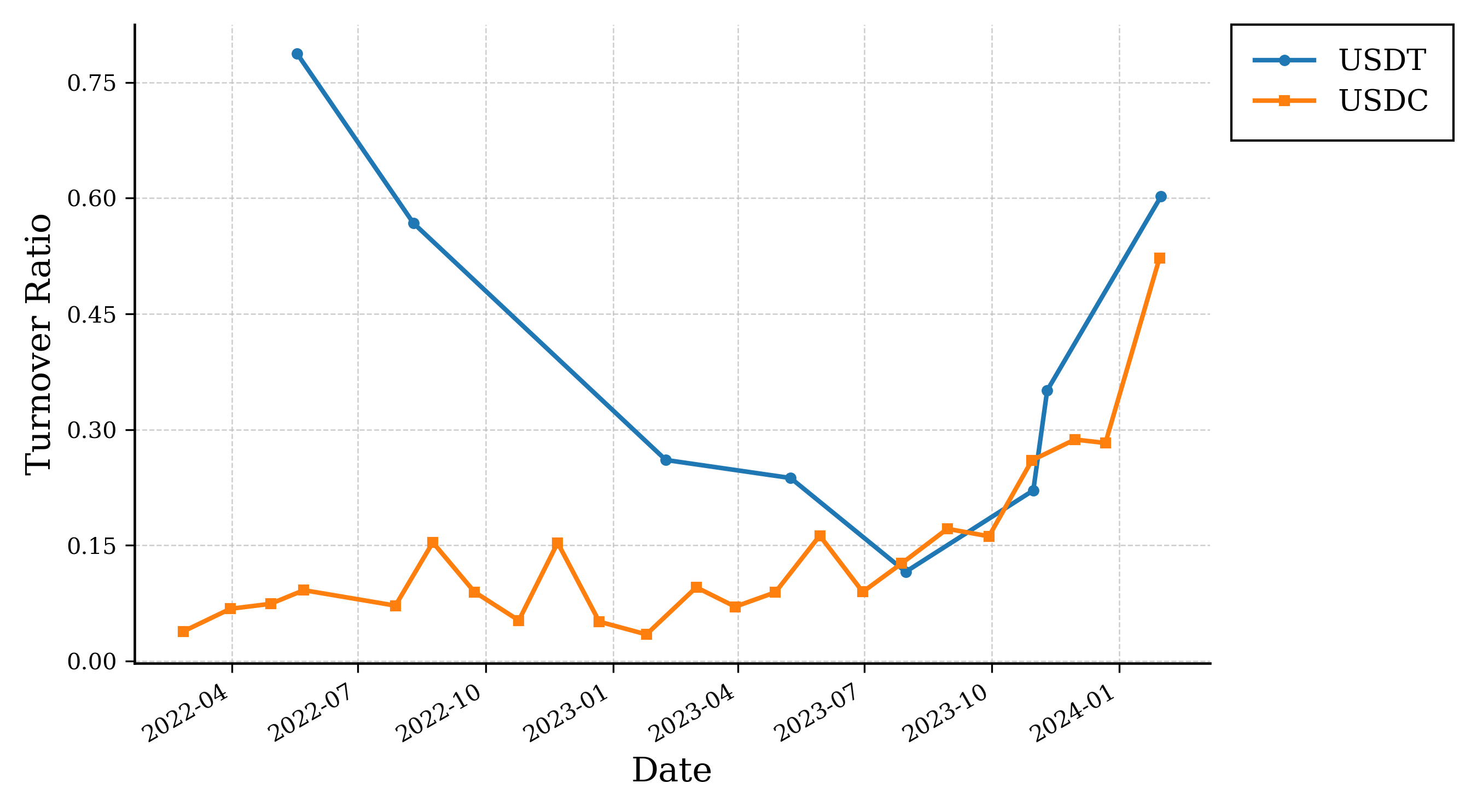}
    \caption{Turnover ratio}
  \end{subfigure}
  \hfill
  \begin{subfigure}[b]{0.48\textwidth}
    \centering
    \includegraphics[width=\textwidth]{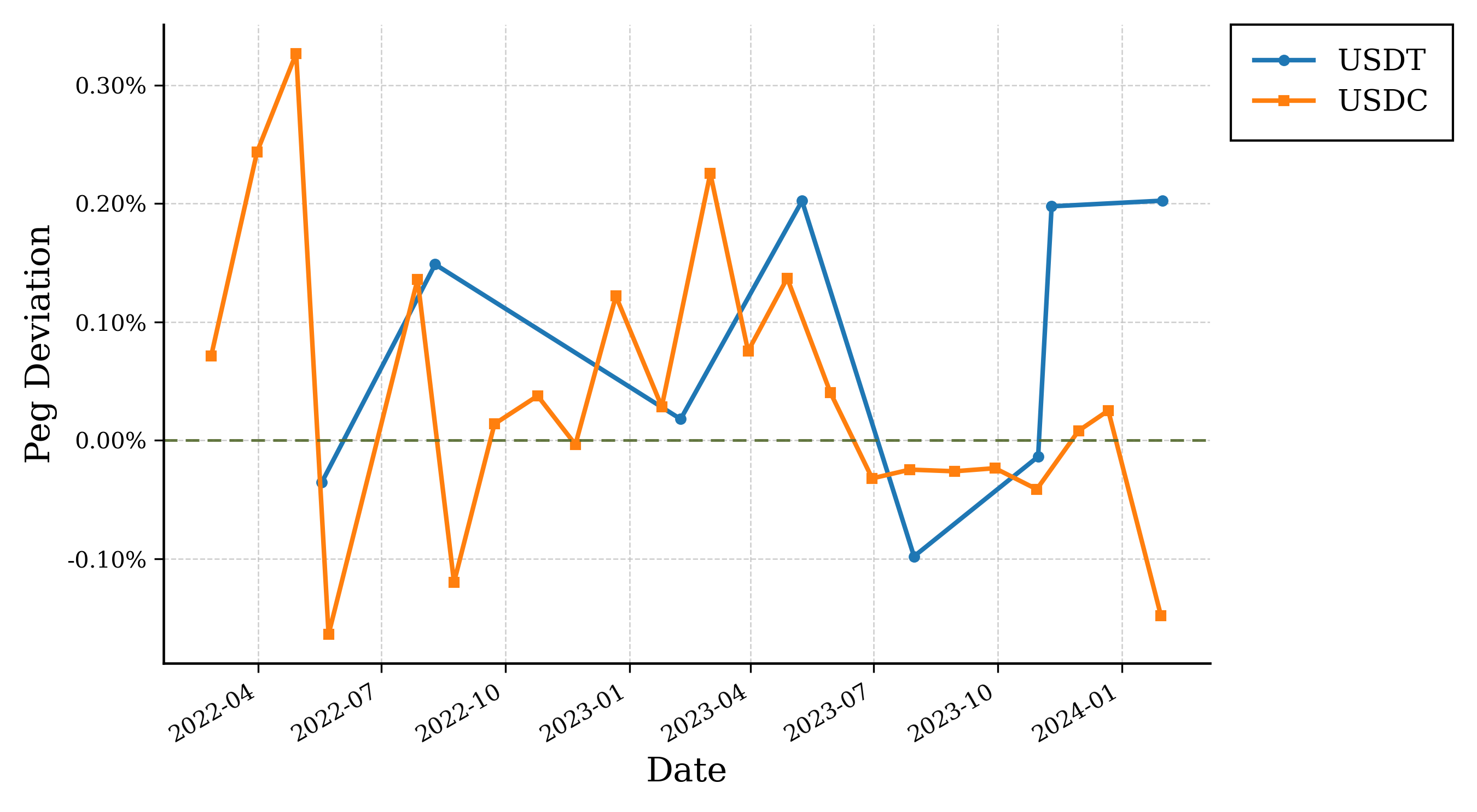}
    \caption{Peg deviation}
  \end{subfigure}
  \caption{Comparative liquidity dynamics and peg stability indicators of major stablecoins.}
  \label{fig:peg_deviation_and_turnover_ratio}
\end{figure}

Third, in terms of \emph{behavioral patterns} and \emph{classification outcomes}, USDT’s few abnormal and suspicious observations align with periods of heightened trading and redemption pressure, implying sensitivity to liquidity shocks and reporting lag. As shown in Fig.~\ref{fig:peg_deviation_and_turnover_ratio}, USDT exhibits sharper swings in peg deviation and a wider range of turnover ratios, often above 0.5 during stress intervals, whereas USDC remains comparatively stable with smaller fluctuations around zero deviation and lower turnover. These patterns reinforce that USDT’s market behavior is more reactive to shocks, while USDC’s steadier profile reflects slower but more controlled adjustment dynamics. By contrast, USDC’s limited suspicious flags are isolated and transitory, reflecting minor operational frictions rather than structural misalignment.

These findings highlight a trade-off between scale-driven liquidity assurance and disclosure-driven transparency assurance. USDT exemplifies a market-dominant but disclosure-light model, where stability depends on liquidity depth and redemption capacity. USDC, although smaller in scale, demonstrates a disclosure-centric equilibrium in which predictable attestation practices and transparent reserve reconciliation underpin market trust. This contrast underscores how stability in fiat-backed stablecoins can emerge from either market liquidity depth or institutionalized transparency, which are two distinct but complementary mechanisms of credibility.

\subsection{Event Analysis}
\noindent\textbf{May 2022: Terra Collapse and USDT Depeg.}
The first stress episode occurred on May 12, 2022, amid the collapse of the algorithmic stablecoin UST between May 7--13, 2022~\cite{de2023intelligent,eichengreen2025stablecoin}. During this period, USDT briefly lost its peg, averaging about \$0.996 with daily trading volume near \$150B. In contrast, USDC remained slightly above parity at around \$1.001 with increased activity. Between May 11 and May 13, USDT’s market capitalization fell by roughly \$3.8B, while USDC’s rose by about \$1.4B, indicating a short-term flight to quality. On May 13, Tether issued an assurance update reaffirming full reserve backing and highlighting a shift toward U.S. Treasury bill holdings to reinforce confidence.

The LLM analysis captured rapid circulation changes and issuer disclosure responses during the episode. Circulation data showed accelerated USDT redemptions and simultaneous USDC inflows into exchanges, consistent with the observed shifts in market capitalization. Issuer disclosures further showed that Tether adjusted its reserve mix within weeks, and the LLM-based multi-agent system identified a temporal alignment between attestation language emphasizing liquidity quality and the observed recovery of peg stability. The model also inferred that disclosure timing lagged the market stress itself, which helps explain why the classification model briefly marked May 2022 as an abnormal disclosure phase.

\smallskip
\noindent\textbf{March 2023: U.S. Banking Turmoil and USDC Depeg.} The second major shock occurred from March 10--13, 2023, when Silicon Valley Bank (SVB) collapsed amid broader U.S. banking turmoil~\cite{wilmarth2023we,diop2024collapse}. \emph{Circle} subsequently disclosed about \$3.3B of USDC reserves held at SVB, triggering a temporary depeg: USDC fell to roughly \$0.88--0.90 on March 11 before rebounding to \$1 by March 13 after regulators guaranteed deposits. During this period, USDT traded at a small premium of about \$1.01 as investors rotated liquidity. Both stablecoins experienced sharp volume increases, but net flows favored USDT, expanding its market share while USDC contracted. 

The LLM detected a pronounced divergence between issuer-reported reserve exposure and observable circulation behavior. It linked the timing of SVB-related disclosures with the observed contraction in USDC mint and burn activity and the rise in USDT transaction velocity. By combining attestation metadata, reserve composition, and transaction-level flow evidence, the LLM suggested that the instability was primarily associated with counterparty concentration rather than redemption mechanics within the token system itself. This interpretation was supported by post-event data, as market equilibrium returned once regulatory assurances restored confidence.

\section{Discussion}
\subsection{Strengths}

The proposed framework demonstrates significant advantages in integrating heterogeneous stablecoin information sources within a unified analytical system. By combining circulation indicators with issuer disclosures under a common MCP architecture, the system enables structured, reproducible, and interpretable analyses that are difficult to achieve through conventional auditing workflows alone. This unified design allows analysts and regulators to compare issuer claims with observable evidence in a timely and consistent manner, improving both accountability and transparency. The use of an LLM such as GPT-5 further enhances interpretability by extracting and normalizing complex attestation narratives, creating a semantic link between human-readable disclosures and quantitative circulation data.

Another key advantage lies in automation and scalability. Manual review of reserve attestations is costly, error-prone, and infrequent. The LLM-driven approach automates this process with high precision and adaptability across issuers and time periods. Its multi-agent structure supports modularity, with each agent responsible for tasks such as disclosure parsing, event detection, or cross-source comparison, allowing new issuers or asset classes to be incorporated with limited overhead. The combination of structured classification and validation mechanisms provides a transparent analytical pipeline while reducing subjectivity and preserving rigor. In practice, this framework can serve as an independent auditing layer for decentralized finance (DeFi), supporting data-driven oversight and the early detection of reserve inconsistencies or systemic risks.

\subsection{Limitations}

Despite its strengths, several limitations remain. First, the reliability of the framework ultimately depends on data quality and availability. Many issuer disclosures are released in unstructured or image-based formats, which can limit extraction accuracy even when using advanced language models. Inconsistent attestation schedules or incomplete reserve details may also hinder precise temporal alignment between disclosures and observable circulation indicators. While the LLM's contextual reasoning mitigates some of these challenges, residual uncertainty remains when disclosed information lacks sufficient detail or consistency. In addition, the framework relies on publicly available sources, so omissions, delays, or selective disclosure can directly constrain the scope and reliability of the analysis.

Second, the framework’s current evaluation logic is partly rule-based, which is interpretable but may not fully capture nonlinear dependencies between market indicators, such as multi-day liquidity feedback or contagion dynamics. The use of threshold-driven classification provides transparency but may oversimplify complex causal relationships during extreme market conditions. Moreover, the model does not yet incorporate real-time streaming data or adaptive calibration based on evolving issuer behavior. Future work should integrate dynamic statistical learning methods, multimodal embeddings, and regulator-verified data pipelines to enhance robustness, precision, and temporal responsiveness.

\section{Conclusion}

This study presents an automated and interpretable framework that employs LLMs to support cross-domain transparency analysis in stablecoins. By integrating circulation data, issuer attestations, and related quantitative indicators within a unified analytical protocol, the framework enables systematic assessment of reserve discrepancies, disclosure timeliness, and stability behavior for leading fiat-backed stablecoins. The empirical analysis of USDT and USDC shows that this approach can surface transparency gaps, align issuer statements with observable evidence, and suggest that market liquidity and disclosure quality represent distinct dimensions of stability. These results highlight the framework’s potential to support more systematic transparency monitoring and stablecoin oversight.


\section*{Acknowledgment}
This research was supported by the OpenAI Researcher Access Program (Project ID: 0000007730).

\bibliographystyle{IEEEtran}
\bibliography{bib}

\end{document}